\documentclass[aps,twocolumn,floatfix,showpacs,groupedaddress]{revtex4}
\usepackage{amsmath,amsfonts,amssymb,graphicx,times,bbm,natbib}
\usepackage[T1]{fontenc}
\usepackage{times}
\def\tp{\tilde{\Phi}}
\def\bc{\frac{\beta_{\rm c}}{2}}
\def\be{\begin{equation}}
\def\ee{\end{equation}}
\begin{document}

\title{Quantum phase transition in the one-dimensional extended
  Peierls-Hubbard model}
\author{H.~Benthien$^1$, F.~H.~L.~Essler$^2$ and A.~Grage$^1$}
\affiliation{$^1$Fachbereich Physik, Philipps-Universit\"at Marburg,
D-35032 Marburg, Germany\\
$^2$The Rudolf Peierls Centre for Theoretical Physics,
University of Oxford, Oxford OX1 3NP, UK} 



\begin{abstract}
We consider the one-dimensional extended Hubbard model 
in the presence of an explicit dimerization $\delta$. For a
sufficiently strong nearest neighbour repulsion we establish the
existence of a quantum phase transition between a mixed bond-order
wave and charge-density wave phase from a pure bond-order wave phase.
This phase transition is in the universality class of the two-dimensional Ising model.
\end{abstract}
\maketitle

\section{Introduction}
It is well known that quasi one-dimensional electron systems
exhibit a ``Peierls instability'' towards the formation of a dimerized
insulating ground state~\cite{peierls}. In the absence of
electron-electron interactions the low temperature phase of such
systems is described in terms of a Peierls insulator with gapped
electron and hole quasiparticle excitations. 
On the other hand, it is well known that undimerized interacting one-dimensional
electron systems are either correlated (Mott) insulators
or Luttinger liquids, see e.g. \cite{GNT,gia}. 
A characteristic feature of these states is that low-lying
excitations are collective modes of charge and spin degrees of freedom 
respectively. An interesting question is then what happens in the case
when strong electron-electron interactions compete with the Peierls
distortion. A simple model having all the necessary ingredients to
study this question is the one-dimensional extended Hubbard model 
with an explicit dimerization $\delta$. Its Hamiltonian is

 \begin{eqnarray}
\hat{H} &=& -t\sum_{l=1,\sigma}^L \left(1+(-1)^{l}\delta\right)
\left(\hat{c}^+_{l,\sigma}\hat{c}_{l+1,\sigma} +\mathrm{h.c.}\right)
\nonumber\\ 
&& +U\sum_{l=1}^L 
\left(\hat{n}_{l,\uparrow}-\frac{1}{2}\right)
\left(\hat{n}_{l,\downarrow}-\frac{1}{2}\right) 
\nonumber\\
&& +V\sum_{l=1}^L 
\left(\hat{n}_{l}-1\right) \left(\hat{n}_{l+1}-1\right)\,,
\label{phm}
\end{eqnarray}
where $\hat{c}_{l,\sigma}^{+}$ creates an electron with spin
$\sigma=\uparrow,\downarrow$  in a Wannier orbital centered around
site $l$, and we have
$\hat{n}_{l,\sigma}=\hat{c}_{l,\sigma}^{+}\hat{c}_{l,\sigma}$,  
$\hat{n}_{l}=\hat{n}_{l,\uparrow}+\hat{n}_{l,\downarrow}$. $U$ is the
on-site and $V$ the next-neighbour Coulomb interaction. Since we are
interested in the half-filled case only, the number of electrons $N$
equals the number of lattice sites $L$. In what follows we will
consider the charge-density wave (CDW) and bond-order wave (BOW) order
parameters 
\begin{eqnarray}
m_{\rm CDW}&=&\frac{1}{L}\sum_{l}
\left(-1\right)^{l}\left(\hat{n}_l-1\right)\ ,\label{eq:cdw-param}\\
m_{\rm BOW} &=& \frac{1}{L}\sum_{l,\sigma}\left(-1\right)^{l}
\left(c_{l,\sigma }^{+}c_{l+1,\sigma }+\textrm{h.c.}\right).
\label{eq:bow-param}
\end{eqnarray}

The model (\ref{phm}) has previously been studied in various
limiting cases. The infinite-$U$ limit was studied in \cite{fg}. The
low-lying excitations in this limit are chargeless spin triplet and
spin singlet excitations, which can be understood in terms of a
spin-Peierls Hamiltonian, see e.g. \cite{spinP,GNT,gia}. The effects
of weak electron-electron interactions in a Peierls insulator were
investigated in \cite{anja} by perturbative methods. Most importantly
for our purposes, the weak-coupling regime $U,|V|\alt t$ was studied
in Refs \cite{tsuchiizu, suzumura}. From the structure of the classical
ground state of the bosonized low-energy effective Hamiltonian
Tsuchiizu and Furusaki showed that increasing the dimerization
$\delta$ from zero for a sufficiently large $V$ drives the system
through a quantum critical point that was argued to be in the
universality class of the two-dimensional Ising model
\cite{orignac}. The mechanism underlying this transition is very
similar to the one exhibited in \cite{FGN1,FGN2}. The purpose of the
present work is to verify the prediction \cite{tsuchiizu}  of an Ising
critical point by means of numerical methods. The outline of the paper
is as follows. In section \ref{ftlimit} we consider the field theory
limit of the model (\ref{phm}) and review and extend the results of
Ref. \cite{tsuchiizu} that suggest the existence of an Ising
transition.  In section \ref{sec:num-res-qpt} we use numerical
techniques to establish that there is indeed an Ising transition in
the lattice model (\ref{phm}). 

\section{Field Theory Limit}
\label{ftlimit}
The weak-coupling limit $U,V \ll t$ of the model (\ref{phm}) is
amenable to a field theory analysis. The low-energy regime can be
described by linearising the non-interacting Fermi spectrum
around the Fermi-points $\pm \pi/2a_0$, where $a_0$ is the lattice
spacing. Applying a bosonization scheme~\cite{GNT,book} in terms of
two Bose fields $\Phi_{\rm c}$ and $\Phi_{\rm s}$ corresponding to
collective charge and spin degrees of freedom respectively we arrive
at the following form of the low-energy Hamiltonian
\cite{tsuchiizu,anja-diss,suzumura}
\begin{equation}
\mathcal{H}=\mathcal{H}_{\rm c}+\mathcal{H}_{\rm s} +\mathcal{H}_{\rm cs}\,,
\label{dsg}
\end{equation}
 where 
\begin{eqnarray}
\mathcal{H}_{\rm c} &=& \frac{v_{\rm c}}{16\pi}\left[\left(
\partial_x\Phi_{\rm c}\right)^2+\left(\partial_x\Theta_{\rm c}\right)^2\right]
-\frac{v_{\rm F}g_{\rm c}}{\pi a_0^2}\cos(\beta_{\rm c}\Phi_{\rm c}),\ 
\label{H1}\\
\mathcal{H}_{\rm s} &=& \frac{v_{\rm s}}{16\pi}\left[\left(
\partial_x\Phi_{\rm s}\right)^2+\left(\partial_x\Theta_{\rm s}\right)^2\right]
\nonumber\\
&+&\frac{v_{\rm F}g_{\rm s}}{\pi a_0^2}\left[\cos(\Phi_{\rm s})+\frac{a_0^2}{16}
\Bigl[(\partial_x\Theta_{\rm s})^2-(\partial_x\Phi_{\rm s})^2\Bigr]\right],
\label{H2}\\
\mathcal{H}_{\rm cs} &=&(4t\delta)/(\pi
a_0)\cos\left(\beta_{\rm c}\Phi_{\rm c}/2\right)
\cos\left(\Phi_{\rm s}/2\right).
\label{H3}
\end{eqnarray}
Here $v_{\rm F}=2ta_0$ is the Fermi velocity of the noninteracting theory
and $\Theta_{c,s}$ are the fields dual to $\Phi_{c,s}$. The bare
values of the couplings $\beta_{c}$, $g_{c,s}$ and charge and spin
velocities $v_{c,s}$ are related to the parameters of the lattice
model as follows (see e.g. \cite{voit})
\begin{eqnarray}
\beta_{\rm c} &=&\left[\frac{2\pi t-V}{2\pi t+U+5V}\right]^\frac{1}{4}
,\\
v_{\rm c} &=& v_{\rm F}\sqrt{\left(1+\frac{U+4V}{4\pi
    t}\right)^2-\left(\frac{U+6V}{4\pi t}\right)^2}\ ,\\
v_{\rm s} &=& v_{\rm F}\left[1-\frac{U}{4\pi t}\right]\ ,\quad
g_{\rm c}=g_{\rm s}=\frac{U-2V}{4\pi t}\ .
\end{eqnarray}
In ${\cal H}_{\rm s}$ we have kept the quadratic term in derivatives in the
interaction part in order to emphasize the SU(2) symmetry of ${\cal
  H}_{\rm s}$. It is of course possible to absorb this term into the
kinetic piece of the Hamiltonian through a rescaling of $\Phi_{\rm
  s}\rightarrow\beta_s\Phi_{\rm s}$, $\Theta_{\rm s}\rightarrow
\beta_s^{-1}\Theta_{\rm s}$. The effective low-energy model consists of two
coupled sine-Gordon models (sGM) and cannot be solved exactly. 
The bosonized expressions for the order parameters are
\begin{eqnarray}
m_{\rm CDW}&\propto& \sin
\left(\bc\Phi_{c}\right)\cos\left(\Phi_{s}/2\right)\ ,\\
\label{eq:cdw-paramFT}
m_{\rm BOW}&\propto& \cos \left(\bc\Phi_{c}\right)\cos
\left(\Phi_{s}/2\right)\,.
\label{eq:bow-paramFT}
\end{eqnarray}
From now on we restrict our analysis to the regime $2V>U$,
which in the absence of dimerization $\delta=0$ corresponds to the
charge-density wave (CDW) regime. In this regime (for $\delta=0$) the
Hamiltonian reduces to the well-known description of the extended
Hubbard model in terms of two sGMs \cite{eduardo}. The charge sector
is described by a sGM with coupling constant $\beta_{\rm c}<1$. As a 
result the charge sector is gapped and the gap scales like  
\begin{equation}
\Delta_{\rm c}\sim t\ |g_{\rm c}|^{1/(2-2\beta_{\rm c}^2)}\ .
\label{chargegap}
\end{equation}
Excitations in the charge sector are scattering states of gapped,
spinless  ``(anti)holons'' carrying charge $\mp e$.
The interaction of spin currents in the spin sector is marginally
relevant and opens up a spectral gap (see e.g. \cite{GNT,eduardo}),
which scales like 
\begin{equation}
\Delta_{\rm s}\sim t\ \exp\bigl(-\frac{1}{2|g_{\rm s}|}\bigr).
\label{spingap}
\end{equation}
The elementary excitations are charge neutral spin-$\frac{1}{2}$
spinon excitations with a spectral gap given by (\ref{spingap}).

\subsection{Classical Ground State}
\label{semclass}
The qualitative behaviour of the field theory (\ref{dsg}) can be
determined by considering the classical limit \cite{FGN1,tsuchiizu}.
The effective potential is given by 
\begin{eqnarray}
\mathcal{U}_{\mathrm{eff}}(\widetilde{\Phi}_{\rm
  c},\widetilde{\Phi}_{\rm s}) &=& \frac{v_{\rm F}}{\pi a_0^2}\Bigl[ 
-g_{\rm c}\cos(\tilde{\Phi}_{\rm c})+g_{\rm s}\cos(\tilde{\Phi}_{\rm
  s})\nonumber\\ 
&& +2 \delta \cos(\tilde{\Phi}_{\rm c}/2)\cos(\tilde{\Phi}_{\rm s}/2)\Bigr],
\label{potentialdichte_cdw}
\end{eqnarray}
where $\tilde{\Phi}_{c}=\beta_{c}\Phi_{c}$,
$\tilde{\Phi}_{s}=\beta_s\Phi_{s}$.
In the charge-density wave regime $U<2V$ we have $g_{\rm c}<0$,
$g_{\rm s}<0$ and $\delta>0$. The structure of the local minima of
${\cal U}_{\rm eff}$ then depends on the value of $\delta$ as follows.

1. In the pure charge-density wave phase $\delta=0$ the minima
are at $\{\tp_{\rm c},\tp_{\rm s}\}= \{(2n+1)\pi,2k\pi\}$, where $n$
and $k$ are integers. The pinning of the fields at these values
implies the presence of a nonzero CDW order parameter $\langle m_{\rm CDW}\rangle\neq 0$. 

2. In the interval $0<\delta<\delta^*=|g_{c}|/(2\pi)$ 
the number of minima is unchanged, but their positions are shifted to 
$\{2\pi(2n+1)\pm \phi_0 , 2\pi(2k)\}$, $\{2\pi(2n) \pm \phi_0 , 2\pi(2k+1)\}$
where $\phi_0=2 \arccos(2\pi\delta/|g_{c}|)$. There are two nonzero
order parameters 
\begin{eqnarray}
\langle m_{\rm CDW}\rangle&\propto&\sqrt{1-(\delta/\delta^*)^2}\ ,\
\langle m_{\rm BOW}\rangle\propto  \frac{\delta
  }{\delta^{*}}\,.\label{pi_semi} 
\end{eqnarray}
For small dimerization $\delta\ll 1$ the CDW order parameter is large
compared to the BOW order parameter and decreases quadratically
$\langle m_{\rm CDW}\rangle\propto
1-\frac{1}{2}\left(\frac{\delta }{\delta^{*}}\right)^{2}$.

3. At a critical dimerization $\delta=\delta^*$ the adjacent 
minima that were moving towards each other merge. Like in the double
sine-Gordon case \cite{mussardo} the analogy with the
$\varphi^4$-description of the Ising model suggests that a quantum phase
transition in the Ising universality class takes place at $\delta^*$.

4. For large dimerizations $\delta>\delta^*$ the positions of
the minima of ${\cal U}_{\rm eff}$ are independent of $\delta$ and
occur at $\{2\pi (2n+1),2\pi (2k)\}$, $\{2\pi (2n),2\pi
(2k+1)\}$. These minima are located at the same positions as in the
pure Peierls insulator. As a result the CDW order parameter now
vanishes, whereas the BOW order parameter stays finite
\begin{equation}
\langle m_{\rm CDW}\rangle= 0,\quad
\langle m_{\rm BOW}\rangle\neq 0.
\end{equation}

\subsection{Perturbation Theory in $\delta$}
\paragraph{Form Factor Perturbation Theory}
\label{ffpt}
In the absence of dimerization $\delta=0$ the field theory
(\ref{H1}), (\ref{H2}) is integrable. Using the knowledge of matrix
elements (form factors) of operators in this integrable theory
\cite{smirnov,FF}, the effects of
$\delta>0$ can be studied by form factor perturbation theory
\cite{FFPT,mussardo}. In the CDW regime $2V>U$ we are dealing with a
fully massive quantum field theory. The changes in the holon and
spinon gaps to first order in $\delta$ are \cite{FFPT}  
\begin{eqnarray}
\Delta_{\rm c}^2(\delta)-\Delta_{\rm c}^2(0)&\sim&
2v_{\rm c}\langle 0|Z_{h}(\theta){\cal H}_{\rm cs}
Z^\dagger_{h}(\theta)|0\rangle\ ,\nonumber\\
\Delta_{\rm s}^2(\delta)-\Delta_{\rm s}^2(0)&\sim&
2v_{\rm s}\langle 0|Z_{s}(\theta){\cal H}_{\rm cs}
Z^\dagger_{s}(\theta)|0\rangle\ .
\label{massshifts}
\end{eqnarray}
Here $Z^\dagger_{\rm s}(\theta)$ is a Faddeev-Zamolodchikov operator
creating a spinon with momentum
$\Delta_{\rm s} \sinh(\theta)/v_{\rm s}$. Similarly $Z^\dagger_h(\theta)$
creates a holon with momentum $\Delta_{\rm c}\sinh(\theta)/v_{\rm c}$. 
The form factors in (\ref{massshifts}) have been calculated in
\cite{smirnov} and substituting them into (\ref{massshifts}) we find
\begin{eqnarray}
\Delta_{\rm c}^2(\delta)-\Delta_{\rm c}^2(0)&\sim&-
\frac{8v_{\rm c}t}{\pi\xi a_0}\delta_{r}\nonumber\\
&&\times\langle0|
\sin\bigl(\bc\Phi_{c}\bigr)\cos\left(\Phi_{s}/2\right)|0\rangle\
,\nonumber\\ 
\Delta_{\rm s}^2(\delta)-\Delta_{\rm s}^2(0)&\sim&{\cal O}(\delta_{r}^2)\ ,
\end{eqnarray}
where $\xi=\beta_{\rm c}^2/(1-\beta_{\rm c}^2)$ and $\delta_{r}$ is the
renormalized coupling constant at a scale set by the gap for
$\delta=0$. The fact that the corrections to the gaps are finite is an
indication that perturbation theory in $\delta$ is well defined at
least for very small values of $\delta$.

\paragraph{Order Parameters for small $\delta$}
Both the classical analysis and the formfactor perturbation theory
considerations suggest that for sufficiently small $\delta$ the
Peierls term can be treated in perturbation theory around the gapped
CDW phase. The same then ought to be the case for the lattice model
itself. Let us then work with the lattice Hamiltonian and represent it 
as $\hat{H}=\hat{H}_0+\delta \hat{H}_1$. The unperturbed extended
Hubbard Hamiltonian $\hat{H}_0$ is invariant under the following
discrete symmetry transformation \cite{book} 
\be
U\hat{c}^+_{j,\sigma}U^\dagger=(-1)^j\hat{c}_{j+1,\sigma}\ ,
\ee
which is a combination of a particle-hole transformation and a
translation by one site. It is straightforward to see that $\hat{H}_1$
is odd under this transformation
\be
U\hat{H}_1U^\dagger=-\hat{H}_1\ .
\label{trafoprop1}
\ee
On the other hand, the order parameters are odd and even respectively
\begin{eqnarray}
U m_{\rm BOW}U^\dagger&=&-m_{\rm BOW}\ ,\nonumber\\
U m_{\rm CDW}U^\dagger&=&m_{\rm CDW}\ .
\label{trafoprop2}
\end{eqnarray}
It follows from (\ref{trafoprop1}) and (\ref{trafoprop2}) that the
perturbative expansions of the order parameters in powers of $\delta$
are of the form
\begin{eqnarray}
\langle m_{\rm CDW}\rangle&=&\sum_{n=0}a_n\delta^{2n}\ ,
\label{eq:orderparam_cdw_perturb}\\
\langle m_{\rm BOW}\rangle&=&\sum_{n=0}b_n\delta^{2n+1}.
\label{eq:pert_bow_order_param}
\end{eqnarray}
The perturbative results for the order parameters,
which obviously can be derived in the field theory limit as well, show
the same kind of dependence on $\delta$ as the one obtained from the
analysis of the classical ground state.

\subsection{Quantum Critical Point}
The 1-loop renormalization group analysis carried out in
Ref.\cite{tsuchiizu} suggests that in the regime we are interested in
the spin degrees of freedom have a ``large'' gap and the low-energy
effective Hamiltonian only involves the charge sector and is given by
a two frequency sGM
\begin{eqnarray}
\mathcal{H}_{c}^{\rm eff} & =&  \frac{v_{c}}{16\pi}
\left[\left(\partial_{x}\Phi_{c}\right)^{2}+\left(\partial
_{x}\Theta_{c}\right)^{2}\right] \nonumber \\
 &-&\frac{v_{\rm F}g_{\rm c}^*}{\pi a_{0}^2}
\cos\left(\beta_{\rm c}^*\Phi_{c}\right)
+\frac{v_Fg_\delta^*}{\pi a_0^2}\cos \bigl(\beta_c^*\Phi_{\rm c}\bigr)\,,
\label{ham_mf_c1}
\end{eqnarray}
where $g_c^*$, $g_\delta^*$ and $\beta_c^*$ are renormalized coupling
constants. It then follows from the analysis of
\cite{mussardo,FGN1,FGN2} that at some critical value of $\delta$ the
charge sector undergoes a quantum phase transition in the universality
class of the two-dimensional Ising model. At the transition point the
charge degrees of freedom become gapless, while the spin degrees of
freedom remain gapped. In the following section we carry out a
numerical analysis of the underlying lattice model in order to assess
the validity of this scenario. 

\section{Numerical Results}
\label{sec:num-res-qpt}
In this section we present numerical results for the CDW- and BOW
order parameters, excitation gaps and critical exponents of the
Hamiltonian (\ref{phm}) obtained with the Density-Matrix
Renormalisation Group (DMRG) \cite{White}. DMRG is known to give
excellent results for ground state expectation values and energies of
one-dimensional lattice Hamiltonians and has become a standard
method in the field. We show that the qualitative results of the
preceding field theoretical analysis are reproduced with DMRG and make
a strong case that the quantum phase transition belongs to the Ising
universality class. We note that the parameters $U$, $V$ chosen in the
numerical analysis are such that the field theory description
(\ref{dsg}) is no longer quantitatively valid as can be seen from the
fact that the numerically determined gaps are no longer small compared
to the bandwidth $4t$, which serves as the cutoff in the field theory
analysis. Our choice of $U$ and $V$ makes the numerical analysis
somewhat easier and shows that the Ising quantum phase transition is a
robust feature of the lattice model.

\subsection{Order Parameters}

\begin{figure}
\includegraphics[width=8.4cm]{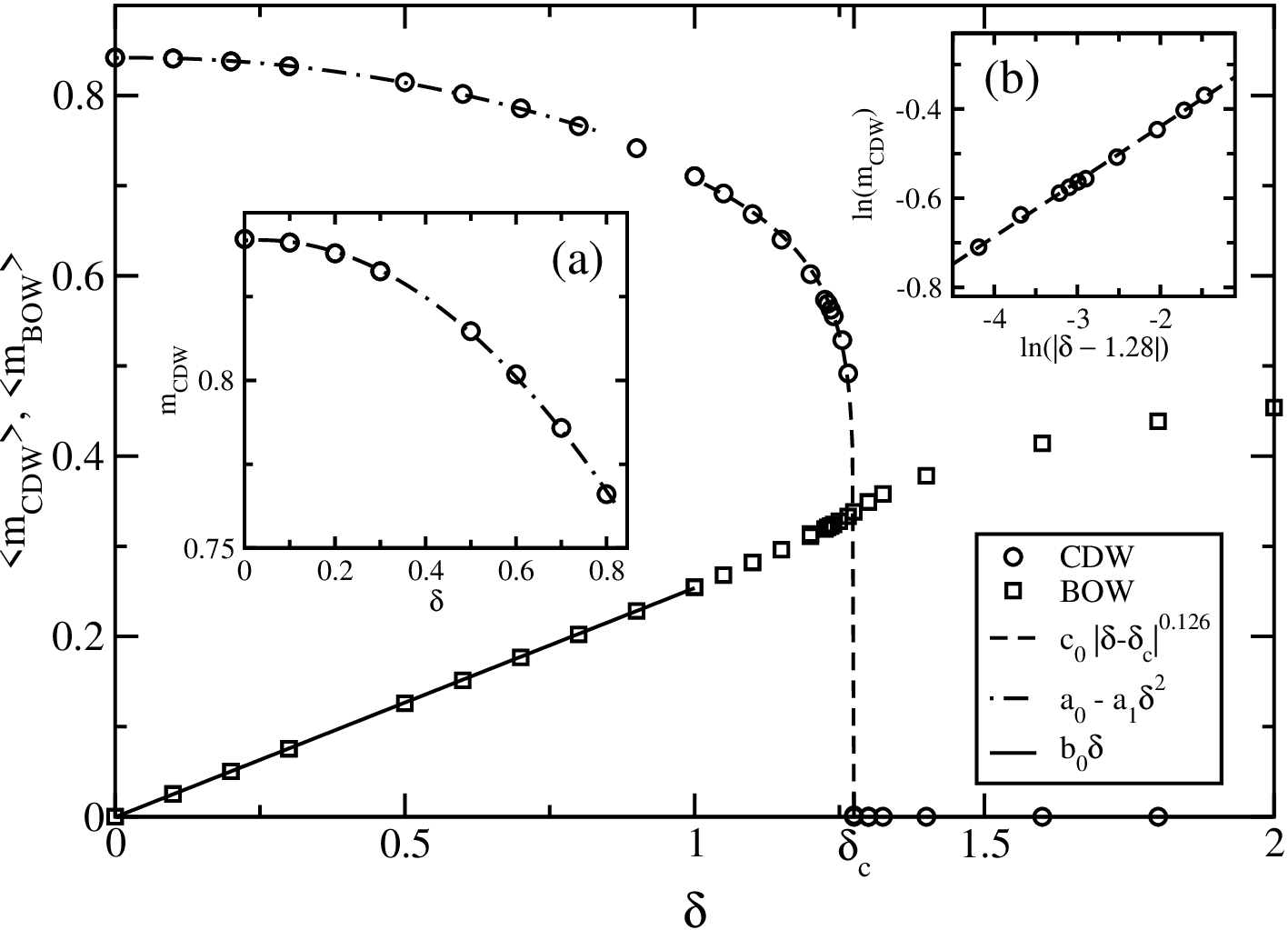}
\caption[CDW and BOW order parameters ($U/t=4$, $V/t=3$)]
{Bond-order wave $\left(\langle m_{{\rm BOW}}\rangle \right)$ and charge-density 
wave~$\left(\langle m_{{\rm CDW}}\rangle\right)$
order parameters of the extended Peierls-Hubbard model with
$U/t=4$, $V/t=3$, and varying dimerization $\delta$. There is a
sharp transition at $\delta_{\rm c}^{\rm OBC}=1.28$. 
For small dimerization $\delta$ clearly $\langle m_{{\rm BOW}}(\delta)\rangle\propto
\delta$ (full line). Inset (a): The CDW order parameter diminishes
quadratically. The dash-dotted line is a quadratic fit of the
form~(\ref{pi_semi}). Inset (b): log-log plot of the CDW order
parameter 
and a fit with the power-law onset 
$\langle m_{\rm CDW}\rangle \sim |\delta-\delta_{\rm c}|^{0.126}$ (dashed line).
}
\label{fig:order-params}
\end{figure}

The DMRG calculations of the order parameters (\ref{eq:cdw-param}),
(\ref{eq:bow-param}) were performed in chains with open boundary
conditions (OBC). We used up to $L=1024$ lattice sites and kept up to
$m=1024$ density-matrix eigenstates in the truncation of the
superblock Hamiltonian. The results are summarized in
Fig.\ref{fig:order-params}.    
At zero dimerization, the system is a CDW-insulator for any value
$V\gtrsim U/2$ \cite{EricEHM}. When we turn on the dimerization,  the 
BOW order parameter $\langle m_{\rm BOW} \rangle $ grows linearly with $\delta$ in
agreement with the semi-classical analysis~(\ref{pi_semi}) and the perturbative
result~(\ref{eq:pert_bow_order_param}).
While $ \langle m_{\rm BOW} \rangle$ is enhanced, the
charge-density wave parameter $\langle m_{\rm CDW} \rangle$ is
reduced until it decays rapidly at the quantum critical point 
\begin{eqnarray}
\delta_{\rm c}^{\rm OBC} =1.28\,. \label{eq:delta-crit-obc}
\end{eqnarray}
Beyond this point $ \langle m_{\rm CDW}\rangle$ vanishes, whereas $\langle m_{\rm BOW}\rangle$
deviates non-trivially from the line $\langle m_{\rm BOW}\rangle = b_0 \delta$ with
$b_0(U/t=4, V/t=3)\approx 0.25$. For small 
values of $\delta$ we can see that $ \langle m_{\rm CDW}\rangle=a_0-a_1\delta^2$ for some
constants $a_0$ and $a_1$. This is in agreement with the predictions
(\ref{eq:orderparam_cdw_perturb}) and (\ref{pi_semi}).
The onset of $\langle m_{\rm CDW}(\delta)\rangle$ as a function of $\delta$ close to
the critical point is strongly reminiscent of the magnetization in the
classical  two-dimensional Ising model. We therefore attempt a fit of
the data with a  power-law onset $\langle m_{\rm CDW}(\delta)\rangle\sim
c_0|\delta-\delta_{\rm c}|^\beta$ in the vicinity of the critical
point and find $\beta=0.126\approx1/8$. The logarithmic plot in the
inset of  figure~\ref{fig:order-params} shows the good agreement of
this fit with our data. 

This confirms our suggestion that the transition belongs to the Ising universality class. A finite-size
scaling analysis of spin and charge excitation gaps in the following subsection further corroborates this 
conclusion.

\subsection{Excitation Gaps}
We define the spin and one-particle gaps
\begin{eqnarray}
\Delta_{\rm s} &=& E_0(N,1) - E_0(N,0)\,, \\
\Delta_{\rm c} &=& E_0(N+1,1/2) + E_0(N-1,1/2)\\
& & -2E_0(N,0)\,, \nonumber
\end{eqnarray}
and the gaps to the first and second excited state
\begin{eqnarray}
\Delta_{\rm 1} &=& E_1(N,0) - E_0(N,0)\,, \\
\Delta_{\rm 2} &=& E_2(N,0) - E_0(N,0)\,.
\end{eqnarray}
In these definitions $E_0(N,S_z)$ is the ground state energy in the subspace
with a given number $N$ of electrons and a given spin $S_z$.  
Likewise, $E_1(N,S_z)$
and $E_2(N,S_z)$ are the energies of the first and second excited state, respectively.

In contrast to the previous section, we do not employ open
boundaries to calculate the excitation gaps since we find that localized bound states occur at the
system boundaries. Since we are not interested in the energy
of such surface states, we  use periodic boundaries (PBC). We studied
periodic chains with an even number of lattice sites and chain lengths
up to $L=128$ while keeping as much as $m=3072$ density-matrix
eigenstates.

\begin{figure}[!ht]
\includegraphics[width=8.6cm]{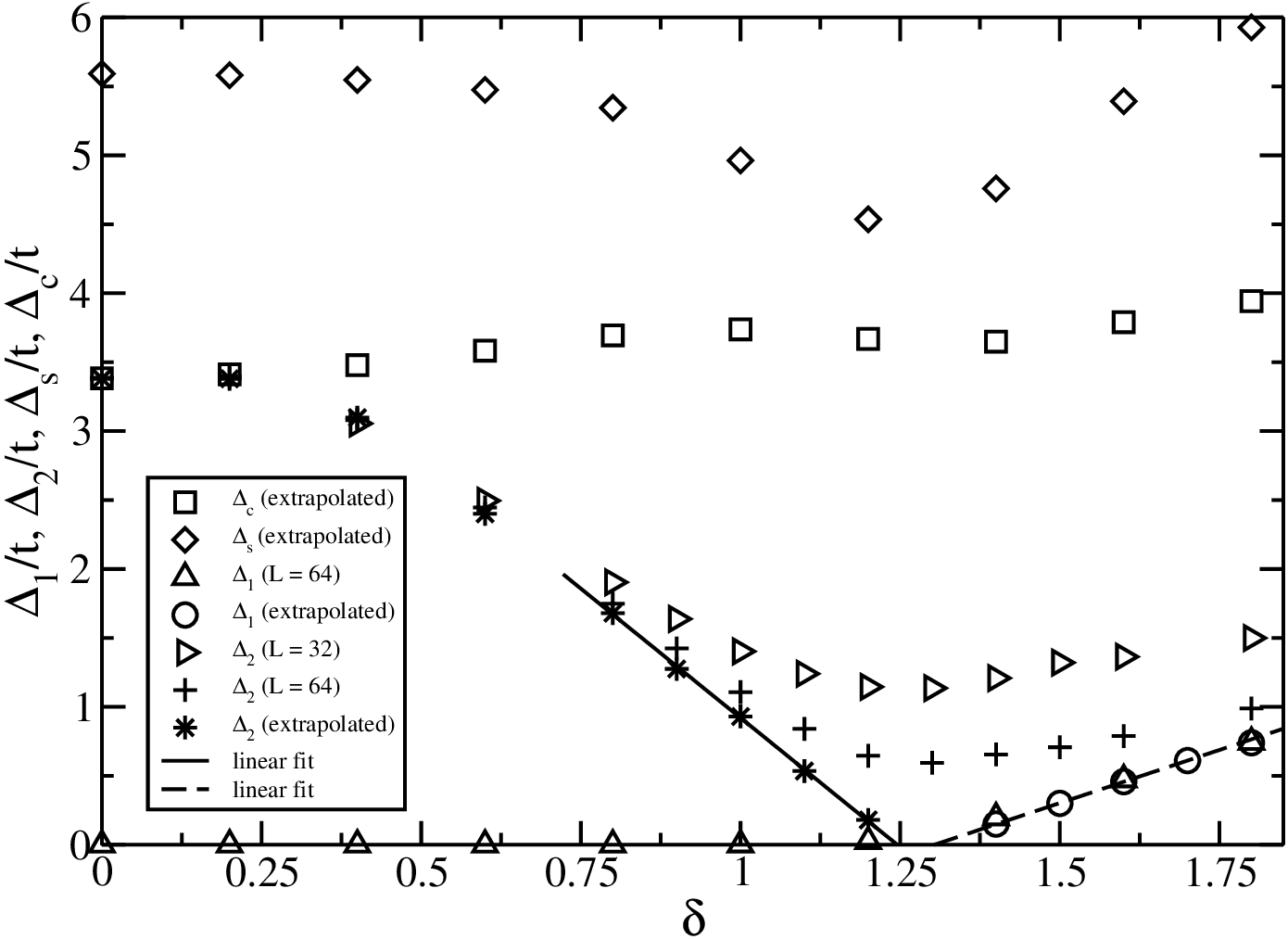}
\caption{{\small Dimerization dependent gaps $\Delta_1$, $\Delta_2$, $\Delta_{\rm s}$, and $\Delta_{\rm c}$.
In the interval $\delta<\delta_{\rm c}$ the ground state is degenerate and
$\Delta_1=0$. The gap $\Delta_2$ is reduced linearly (full line) as we approach $\delta_{\rm c}$.
Above the transition the gap to the first excited state, $\Delta_1$, opens linearly (dashed line).
The spin gap
$\Delta_{\rm s}$ is non-zero for all $\delta$ and is equal to $\Delta_2$ for very
small dimerization. We always have $\Delta_{\rm c}(\delta)>\Delta_{\rm s}(\delta)$ for the
one-particle gap $\Delta_{\rm c}$.}}
\label{fig:gaps-ephm}
\end{figure}

Figure~\ref{fig:gaps-ephm} shows a plot of the gaps as a
function of the dimerization. The gaps $\Delta_1$ and $\Delta_2$ are strongly size dependent. 
Therefore,
we apply finite-size scaling analysis to extrapolate the gaps to the thermodynamic limit.
Below the critical dimerization the gap $\Delta_1$ extrapolates to values very close to zero. 
This means that the ground state is twofold degenerate in the CDW phase.
Above the critical dimerization the gap $\Delta_1$ opens linearly and the ground
state is no longer degenerate and displays no long-range CDW order. 
At $\delta=0$ the extrapolated gap to the second excited state 
$\Delta_2$  is very close to the value of the spin gap $\Delta_{\rm s}$ which we expect to
be equal in a CDW-insulator. They stay close also for small dimerizations which indicates that the
CDW phase of the extended Hubbard model is not strongly perturbed by a small dimerization.
Tuning
$\delta$ to larger values the spin gap $\Delta_{\rm s}$ is not much affected in contrast to
$\Delta_2$ which is now linearly reduced with growing dimerization.
Above the transition $\Delta_2$ increases with the dimerization. Figure~\ref{fig:gaps-ephm} suggests
that $\Delta_2$  is at most slightly larger than $\Delta_1$ in the thermodynamic limit or possibly 
degenerate. We find that the one-particle gap $\Delta_{\rm c}>\Delta_{\rm s}$ for any dimerization.
 
\begin{figure}
\includegraphics[width=8.6cm]{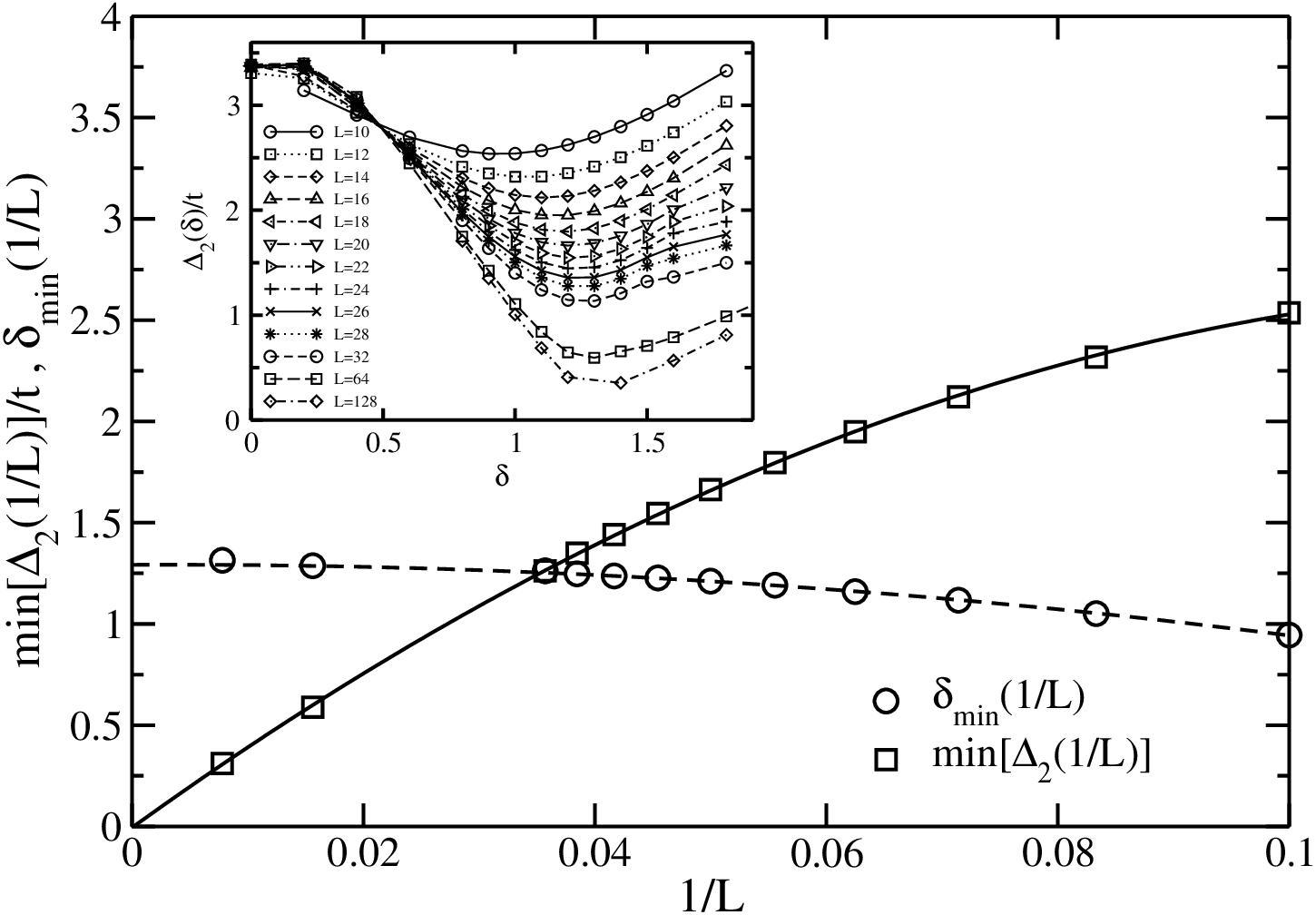}
\caption{Extrapolation of the minima $\min[\Delta_2(\delta)]$ and their positions $\delta_{\rm min}$
to the limit $1/L\to0$ for periodic boundary conditions. Within the numerical precision
$\min[\Delta_2(1/L\to0)]=0$ and $\delta_{\rm c}^{\rm PBC}=1.29$. Inset:
$\Delta_2(\delta)$ for various $L$.}
\label{fig:delta-extrap}
\end{figure}

In order to determine the critical point for the periodic chains we calculated the dimerization
dependence of $\Delta_2$ for many different system sizes. The results are shown in the inset of 
figure~\ref{fig:delta-extrap}.
We determine the minima $\min[\Delta_2(\delta)]$ and their positions $\delta_{\rm min}(L)$ by
fitting second order polynomials to the curves $\Delta_2(\delta)$ for various system lengths $L$.
We then extrapolate these quantities to the thermodynamic limit. This is shown in
figure~\ref{fig:delta-extrap} where we observe that $\Delta_2(\delta_{\rm min})\to0$ within the
precision of our extrapolation. 
The value of the critical dimerization determined from $\delta_{\rm min}(1/L=0)$ reads
\begin{equation}
\delta_{\rm c}^{\rm PBC}=1.29\,,\label{eq:delta-crit-pbc}
\end{equation}
in good agreement with the result (\ref{eq:delta-crit-obc}) previously obtained with open
boundaries.

We have seen that the onset of $\langle m_{\rm CDW}\rangle$ is compatible with an
Ising-type phase transition. Now, we can go further to show that the
excitation of the system that becomes critical at $\delta_{\rm c}$
also suggests this interpretation. As $\delta$ approaches $\delta_{\rm
  c}$ we expect that the gap to the lowest excitation vanishes like \cite{sachdev}
\begin{equation}
\Delta^{\pm} \sim A^{\pm} |\delta - \delta_{\rm c}|^{z \nu}\,, \label{eq:vanishing-gap}
\end{equation}
below ($-$) and above ($+$) the critical point. The non-universal
constant $A^{\pm}$ is a typical energy scale of the system and $z \nu$
is a universal critical exponent.  In figure~\ref{fig:gaps-ephm} we
show a linear fit of the extrapolated gap to the lowest excited state
above ($\Delta^+=\Delta_1$) and below ($\Delta^-=\Delta_2)$ the
transition point. Both fits are a good description of our data since
we can derive a critical dimerization $\delta_{\rm c}\approx 1.25$
which is consistent with (\ref{eq:delta-crit-obc}) and (\ref{eq:delta-crit-pbc}).
We can now infer that $z\nu = 1$. In order to fix $z$ we note that the
characteristic length scale $\xi(\delta)$ of the critical fluctuations
diverges at the critical point such that $\xi \sim |\delta-\delta_{\rm
  c}|^{-\nu}$ holds. The length scale $\xi(\delta)$ can be estimated
by considering the critical dimerization $\delta_{\rm c}(L)$ as a
function of the system length $L$. By inverting this relation we
obtain a critical system size $L_{\rm c}(\delta)$ which is an estimate
of the length scale $\xi(\delta)$ of the critical fluctuations. From
this we find that $\nu=0.98\approx1$. Since the characteristic energy
scale $\Delta^{\pm}$ vanishes linearly we conclude that the dynamical
critical exponent $z=1$. Since both  $\beta\approx 1/8$ and
$\nu\approx1$ are independent universal exponents we may conclude that
the observed quantum phase transition belongs to the Ising universality class.

\section{Conclusions}
We have shown that there is a quantum phase
transition from a mixed CDW-BOW to a BOW phase in the half-filled
extended Peierls-Hubbard model. Field theory arguments suggest that
this phase transition belongs to the universality class of the
two-dimensional Ising model.  A DMRG study of the extended
Peierls-Hubbard model for parameters $U/t=4$, $V/t=3$ reveals that 
there is a transition at a critical value $\delta_{\rm c} \approx 1.3$,
where the CDW order parameter 
$\langle m_{\rm CDW} \rangle$ is found to vanish. A detailed analysis of the order
parameters and excitation gaps in the vicinity of the transition
confirms that the transition falls into the Ising universality class.
This is a robust property of the lattice model away from the 
weak-coupling limit.

\acknowledgments
We are grateful to F.~Gebhard, A. A. Nersesyan and A. M.~Tsvelik for
important discussions. HB and AG acknowledge support by the Theory
Institute for Strongly Correlated and Complex Systems at BNL and the
Optodynamics Center of the Universit\"at Marburg. 
FHLE acknowledges support by the EPSRC under Grant GR/R83712/01 and
thanks the ICTP at Trieste, where this work was completed, for
hospitality. 


\end{document}